\documentclass[prl,superscriptaddress, preprintnumbers,twocolumn,showpacs]{revtex4}
\usepackage{amsmath}
\usepackage{amssymb}
\usepackage{amsthm}
\usepackage{amsfonts}
\usepackage{algorithmic}
\usepackage{enumerate}
\usepackage{latexsym}
\usepackage{graphicx}
\usepackage{bm}
\usepackage{subfigure}
\usepackage{color}

\begin{document}

\title{Interplay between edge and bulk states in silicene nanoribbon}

\author{Xing-Tao An}
\email{anxingtao@semi.ac.cn} \affiliation{School of Sciences, Hebei University of
Science and Technology, Shijiazhuang, Hebei 050018, China}\affiliation{SKLSM, Institute of
Semiconductors, Chinese Academy of Sciences, P. O. Box 912, Beijing
100083, China}
\author{Yan-Yang Zhang}
\affiliation{SKLSM, Institute of Semiconductors, Chinese Academy of
Sciences, P. O. Box 912, Beijing 100083, China}
\affiliation{International Center for Quantum Materials, Peking
University, Beijing 100871, China}
\author{Jian-Jun Liu}
\affiliation{Physics Department, Shijiazhuang University,
Shijiazhuang 050035, China}
\author{Shu-Shen Li}
\affiliation{SKLSM, Institute of Semiconductors, Chinese Academy of
Sciences, P. O. Box 912, Beijing 100083, China}

\date{\today}

\begin{abstract}
We investigate the interplay between the edge and bulk states,
induced by the Rashba spin-orbit coupling, in a zigzag silicene
nanoribbon in the presence of an external electric field. The
interplay can be divided into two kinds, one is the interplay
between the edge and bulk states with opposite velocities, and the
other is that with the same velocity direction. The former can open
small direct spin-dependent subgaps. A spin-polarized current can be
generated in the nanoribbon as the Fermi energy is in the subgaps.
While the later can give rise to the spin precession in the
nanoribbon. Therefore, the zigzag silicene nanoribbon can be used as
an efficient spin filter and spin modulation device.
\end{abstract}

\maketitle

Graphene, a two-dimensional honeycomb network of carbon atoms, has
received great attention in recent years due to its unique physical
properties and application potentials in future nanoelectronic
devices.\cite{Novoselov, Geim, Sarma} The discovery of graphene
forms an platform to explore the properties of two-dimensional
honeycomb electronic systems. However, its compatibility with
current silicon-based nanotechnologies may face challenges.
Recently, strong effort has been invested to search theoretically
and experimentally for two-dimensional honeycomb structures formed
by other elements, such as silicon.\cite{Padova, Houssa, Lalmi,
CCLiu, Vogt, Drummond, Ezawa1, Ezawa2, Kang, An1, CCLiu2, Tahir,
An2, Chen} Silicene, a sheet of silicon atoms forming a honeycomb
lattice analogous to graphene, has been theoretically predicted and
successfully synthesized.\cite{Padova, Houssa, Lalmi, CCLiu, Vogt}
Many striking electric properties of graphene, such as zero gap,
linear dispersion of the electron band and high Fermi velocity,
could be transferred to silicene.

Different from graphene, silicene has a buckled structure owing to a
large ionic radius of silicon, which creates new possibilities for
manipulating the dispersion of electrons and controlling band gap
electrically in silicene.\cite{Drummond} Furthermore, silicene has a
relatively large spin-orbit gap of $1.55meV$ that may induce quantum
spin Hall effect and quantum anomalous Hall effect.\cite{CCLiu,
Ezawa2} A topological phase transition from a quantum spin Hall
state to a band insulator can be induced by an external electric
field in silicene.\cite{Ezawa1} A valley polarized quantum Hall
effect has also been demonstrated in the presence of a perpendicular
external magnetic field.\cite{Ezawa2} The extraordinary transport
properties of silicene nanoribbons have also been studied
theoretically. For example, Kang et al., using first principles
calculations, studied the symmetry-dependent transport properties
and magnetoresistance effect in silicene nanoribbons.\cite{Kang} The
spin-polarized current induced by a local exchange field in a
silicene nanoribbon has been investigated in our previous
work.\cite{An1}

In this Letter, we study the interplay between the edge and bulk states
induced by the Rashba spin-orbit coupling in the zigzag silicene
nanoribbon in the presence of an external electric field, by using
the nonequilibrium Green's function method. Due to the effective
spin-orbit coupling and the staggered sublattice potential induced
by an external electric field, the spin polarization in silicene is
opposite at different valleys, which is called the valley-spin
locking.\cite{Ezawa1} The interplay between the edge and bulk states
occurs when the Rashba spin-orbit coupling exists in the silicene
nanoribbon. Different from subband mixing due to the spin-orbit coupling
in the conventional semiconductor quantum wires,\cite{Wang, Moroz,
Mireles, Wang2, Su, Sun, Zhang} the interplay between the edge and bulk
states with opposite group velocity opens a small, direct and
spin-dependent subgap. At a given Fermi level in this subgap,
an obvious spin polarized current can be obtained because of
the disequilibrium of spin-up and spin-down states with positive
group velocities (i.e., right going channels).

The silicene system with an external electric field is described by the the following Hamiltonian:\cite{CCLiu2}
\begin{eqnarray}
H&=&-t\sum_{\langle{ij}\rangle\alpha}c_{i\alpha}^{\dag}c_{j\alpha}+i\frac{\lambda_{SO}}{3\sqrt{3}}\sum_{\langle\langle{ij}\rangle\rangle\alpha\beta}\nu_{ij}c_{i\alpha}^{\dag}\sigma_{\alpha\beta}^{z}c_{j\beta}\nonumber\\
&-&i\frac{2}{3}\lambda_{R}\sum_{\langle\langle{ij}\rangle\rangle\alpha\beta}\mu_{i}c_{i\alpha}^{\dag}({\bm \sigma}\times{\bm d}_{ij}^{0})^{z}_{\alpha\beta}c_{j\beta}\nonumber\\
&+&\lambda_{\nu}\sum_{i\alpha}\mu_{i}c_{i\alpha}^{\dag}c_{i\alpha}\label{eq1},
\end{eqnarray}
where $c_{i\alpha}^{\dag}$ creates an electron with spin
polarization $\alpha$ at site $i$; $\langle{ij}\rangle$ and
$\langle\langle{ij}\rangle\rangle$ run over all the nearest and
next-nearest neighbor hopping sites, respectively. The first term is
the nearest-neighbor hopping with the transfer energy $t=1.6eV$. The
second term describes the effective spin-orbit coupling, where
${\bm \sigma}=(\sigma_{x}, \sigma_{y}, \sigma_{z})$ is the Pauli
matrix of spin and $\nu_{ij}$ is defined as
$\nu_{ij}=({\bm d}_{i}\times{\bm d}_{j})/|{\bm d}_{i}\times{\bm d}_{j}|=\pm{1}$
with ${\bm d}_{i}$ and ${\bm d}_{j}$ the two bonds connecting the
next-nearest neighbors ${\bm d}_{ij}$. The third term represents the
Rashba spin-orbit coupling, where $\mu_{i}=\pm1$ for the A (B) site,
and ${\bm d}^{0}_{ij}={\bm d}_{ij}/|{\bm d}_{ij}|$. The fourth term
is a staggered sublattice potential and its strength $\lambda_{\nu}$
can be tuned by a perpendicular electric field due to the buckling
distance between two sublattices of the silicene.

\begin{figure}[htb]
\includegraphics[scale=0.5,angle=0]{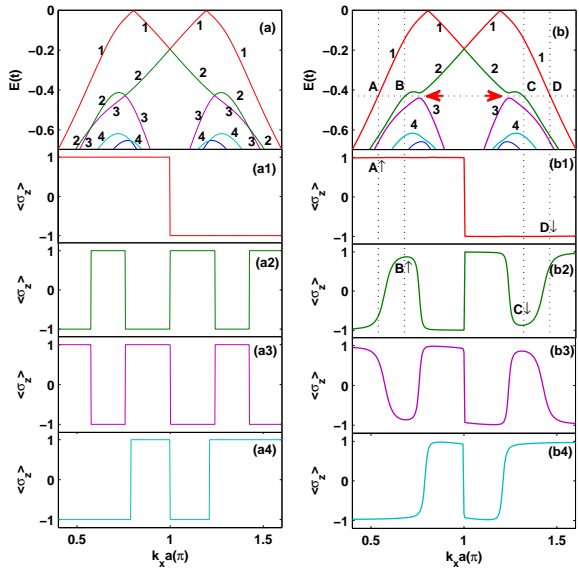}
\centering \caption{(color online) The energy spectrum (a), (b) and the corresponding spectrum of the spin $<\sigma_{z}>$ (a1-a4) and (b1-b4) of a zigzag silicene nanoribbon for $\lambda_{R}=0$ (a) and $\lambda_{R}=0.07t$ (b). The numbers in (a) and (b) denote the different states. At a given Fermi level in the subgap there exist four different states, which are labeled as A, B, C, and D, in which $\uparrow$ ($\downarrow$)stands for the up (down) spin polarization. The horizontal arrows in (b) point to the small subgaps in the energy spectrum.}
\label{figone}
\end{figure}

In order to assure the system in the quantum spin Hall state, we set
the parameters $\lambda_{SO}=0.3t$ and
$\lambda_{\nu}=0.2t$.\cite{Ezawa1,Ezawa2} The length of the zigzag
silicene nanoribbon $L$ is taken to be infinite. The energy spectrum
of the zigzag silicene nanoribbon, together with the corresponding
eigenfunctions $\varphi_{n}(k_{x})$, can be numerically obtained by
diagonalizing the Hamiltonian for each momentum $k_{x}$ in the $x$
direction.\cite{LiH} The calculated energy spectrum of the zigzag
silicene nanoribbon with width $W=14a/\sqrt{3}$ ($N_{y}=20$), where
$a$ is the next nearest-neighbor distance, for $\lambda_{R}=0$ and
$\lambda_{R}=0.07t$ is plotted in Figs. 1(a) and (b), respectively.
Edge states appear in the bulk band gap of the energy spectrum
whether there exists Rashba spin-orbit coupling or not, since this
small $\lambda_{R}$ does not lead to a topological phase transition.

Let's focus on the top of the bulk valence band, where it meets the
edge states, i.e., subbands labeled as 2 and 3 in Figs. 1(a) and
(b). Without Rashba spin-orbit coupling (Fig. 1(a)), the edge states
cross with the bulk valley with opposite spin,\cite{Ezawa2} as can
be seen in Figs. 1 (a2) and (a3). In this case, there is no
interplay between them due to the absence of spin-flip effects.
Finite $\lambda_{R}$, on the other hand, couples these two subbands
and leads to anti-crossings between them as seen in Fig. 1(b). There
are two cases for interplays between the edge and bulk subbands
(channels). The first case (anti-parallel crossing) happens at the
energy $E\sim -0.43t$ near the Dirac point, between the states with
opposite group velocities $v$
($v=1/\hbar(\partial{E}/\partial{k_{x}})$). The second case
(parallel crossing), however, corresponds to the crossing between
states with the same direction of velocity, at the energy
$E\sim-0.6t$. Notice that the subgap opened by the Rashba spin-orbit
coupling is direct (indirect) for the anti-parallel (parallel)
crossing, respectively. This is the essential physics we will rely
on in this work.

We also calculate the $k_{x}$-dependent expectation value of spin,
$\langle\sigma_{z}\rangle=\langle\varphi_{n}(k_{x})|\sigma_{z}|\varphi_{n}(k_{x})\rangle$
for the $n$th occupied states. The calculated
$\langle\sigma_{z}\rangle$ for the same parameters as in Figs. 1(a)
and 1(b) are shown in Figs. 1(a1-a4) and (b1-b4), in which the color
of the line denotes different states. We can see from Figs. 1(a1) and
(b1) that the Dirac cone around K (K') point is polarized with spin
up (down) due to the effective spin-orbit coupling and the external
electric field for the system with or without Rashba spin-orbit
coupling. If $\lambda_{R}=0$, $\sigma_{z}$ and the Hamiltonian $H$
commute. Therefore, $\sigma_{z}$ is a good quantum number, and the
expectation value of spin $\langle\sigma_{z}\rangle$ consists of
just two values $\pm1$, as shown in Figs. 1 (a1-a4). When the Rashba
term is turned on, $\sigma_{z}$ no longer commutes with the
Hamiltonian $H$. In this case, a continuous variation between $+1$
and $-1$ in spin polarization of the states with $k_{x}$ appear.

At a given Fermi level in the direct subgap opened by the
Rashba spin-orbit coupling, there exist four different states
labeled as A, B, C, and D. From Fig. 1(b), one can easily find that
the electrons in states A and B are right-going waves with velocity
in the positive $x$ direction, while the electrons in states C and D
are left-going waves with velocity in the negative $x$ direction. We
can also examine the spin polarization of the states from Figs. 1(b1)
and (b2), states A and B being almost fully spin-up polarized and
states C and D spin-down polarized. Therefore, under a definite
arrangement of bias voltage, there remains only the right-going
channel with spin-up electrons, that is, a spin-polarized current
appears in the system as the Fermi level is in the direct subgaps opened by
the Rashba spin-orbit coupling.

\begin{figure}[htb]
\centering
\includegraphics[scale=0.5,angle=0]{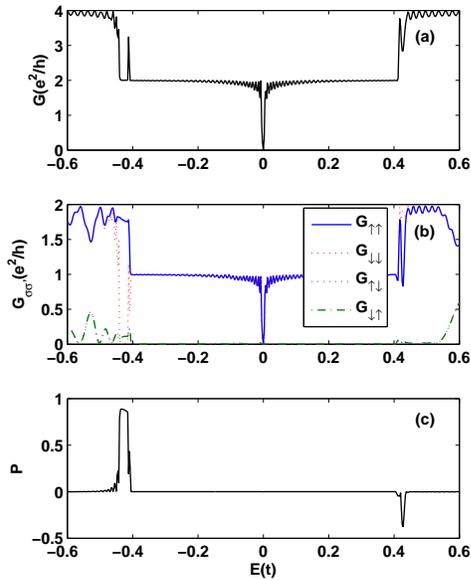}
\caption{(Color online) The total conductance $G$ (a), spin-dependent conductance $G_{\sigma\sigma'}$ (b) and spin polarization $P$ (c) of the zigzag silicene nanoribbon vs $E$ for $\lambda_{R}=0.07t$.} \label{figtwo}
\end{figure}

In order to investigate this spin polarization in a more direct way,
we calculate the spin-dependent conductance and spin polarization
using the nonequilibrium Green's function method. The following
discussion is based on the assumption that only the nearest-neighbor
hopping exists in the left and right leads, i.e., the Hamiltonian of
lead-$p$ is simply
$H_{p}=-t\sum_{\langle{ij}\rangle\alpha}c_{i\alpha}^{\dag}c_{j\alpha}$.
The conductance matrix is calculated by Landauer
formula\cite{Kariminezhad}
\begin{eqnarray}
G=\left(
\begin{array}{ccc}
    G_{\uparrow\uparrow} & G_{\uparrow\downarrow}\\
    G_{\downarrow\uparrow} & G_{\downarrow\downarrow}\\
\end{array}
\right)
=\frac{e^{2}}{h}\sum_{i,j=1}^{N_{y}}\left(
\begin{array}{ccc}
    |t_{ij,\uparrow\uparrow}|^{2} & |t_{ij,\uparrow\downarrow}|^{2}\\
    |t_{ij,\downarrow\uparrow}|^{2} & |t_{ij,\downarrow\downarrow}|^{2}\\
\end{array}
\right).
\end{eqnarray}
The transmission matrix is $\mathrm{t}=2\sqrt{\Gamma_{L}}G_{1N_{x}}^{r}\sqrt{\Gamma_{R}}$, where $\mathbf{\Gamma}_{p}(E)=i[\mathbf{\Sigma}_{p}^{r}(E)-\mathbf{\Sigma}_{p}^{a}(E)]$ is the line-width function with a well-defined matrix square root and $G_{1N_{x}}^{r}$ that connects the unit cells $1$ and $N_{x}$ along the direction of transport is the $4N_{y}\times4N_{y}$ submatrix of the full Green function matrix. The retarded (advanced) self-energy $\mathbf{\Sigma}_{p}^{r}$ ($\mathbf{\Sigma}_{p}^{a}=\left[\mathbf{\Sigma}_{p}^{r}\right]^{\dag}$) describing the interaction of the sample with the lead-$p$ can be calculated numerically.\cite{Sancho} The total conductance $G$ and the spin polarization $P$ in lead-$R$ can be respectively defined as
\begin{eqnarray}
G=G_{\uparrow\uparrow}+G_{\downarrow\uparrow}+G_{\uparrow\downarrow}+G_{\downarrow\downarrow}
\end{eqnarray}
and
\begin{eqnarray}
P=\frac{G_{\uparrow\uparrow}+G_{\downarrow\uparrow}-G_{\uparrow\downarrow}-G_{\downarrow\downarrow}}{G_{\uparrow\uparrow}+G_{\downarrow\uparrow}+G_{\uparrow\downarrow}+G_{\downarrow\downarrow}}.
\end{eqnarray}

We assume that the length of the zigzag silicene nanoribbon is
$L=100a$ ($N_{x}$=100). Figs. 2(a), (b) and (c) show the total
conductance, spin-dependent conductance and spin polarization versus
energy $E$ for $\lambda_{R}=0.07t$, respectively. In Fig. 2(a), the
plateau like structure of the total conductance in units of
$2e^{2}/h$ is observed immediately. However, the resonance like
structure is superimposed on the conductance plateau because of the
mismatch between the central sample and the leads. Due to the finite
size effect, the edge states of the sample are coupled with each
other and a small gap in the energy spectrum of the edge states is
opened, so a narrow dip emerges in the conductance at $E=0$. From
Fig. 2(b) we can find that the spin-up and spin-down electrons are
not mixed i.e., $G_{\uparrow\downarrow}=G_{\downarrow\uparrow}=0$
when the energy $E$ is in the bulk gap because there are only edge
states protected by topological invariants in the bulk gap. When the
energy $E$ is in the bulk band, the interplay between the edge and
bulk states induced by the Rashba spin-orbit coupling can lead to
the spin precession. Therefore in this case,
$G_{\uparrow\downarrow}=G_{\downarrow\uparrow}\neq0$. It is more
interesting that as the energy $E$ is in the direct subgaps opened by the
Rashba spin-orbit coupling, $G_{\uparrow\uparrow}$ is very large
while $G_{\downarrow\downarrow}$ is almost equal to zero because the
electrons traveling from left to right in the zigzag silicene
nanoribbon are almost spin up. Therefore, the spin polarization can
be very large when the energy is in the direct subgaps opened by the Rashba
spin-orbit coupling, as shown in Fig. 2(c). The spin polarization can
be also obtained at $E\sim0.43t$ due to the interplay between the
edge and conduction subbands. However, the value of the spin
polarization is smaller than that at $E\sim-0.43t$ because the
electron-hole symmetry is broken by the Rashba spin-orbit coupling.

\begin{figure}[htb]
\centering
\includegraphics[scale=0.5,angle=0]{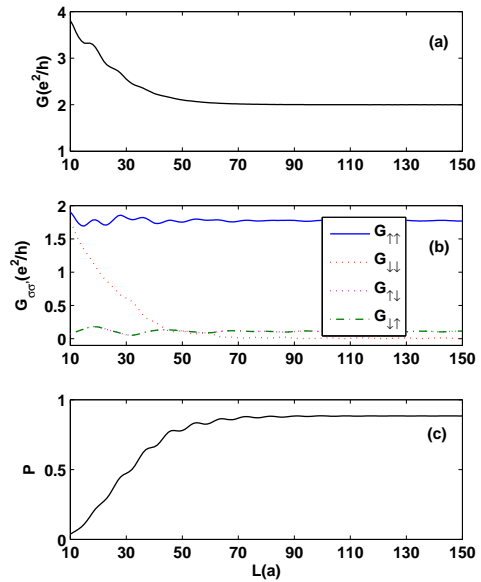}
\caption{(Color online) The total conductance $G$ (a), spin-dependent conductance $G_{\sigma\sigma'}$ (b) and spin polarization $P$ (c) vs the length $L$ of the zigzag silicene nanoribbon for $\lambda_{R}=0.07t$ and $E=-0.43t$.} \label{figthree}
\end{figure}

In Fig. 3, we show the total conductance $G$, spin-dependent
conductance $G_{\sigma\sigma'}$ and spin polarization $P$ versus the
length of the zigzag silicene nanoribbon $L$ for $E=-0.43t$ which is
in the direct subgaps opened by the Rashba spin-orbit coupling. This
case corresponds to the interplay between the traveling edge and
bulk states with the opposite velocity. When the length of the
nanoribbon is very short, for example $L=10a$, the band structure as
shown in Fig. 1(b) has not been formed well and the electrons
quickly tunnel through the sample. As the length of the nanoribbon
increases to $50a$, the total conductance becomes $2e^{2}/h$ and the
conductance $G_{\downarrow\downarrow}$ becomes zero because there
are only two spin-up modes involved in transport and the spin-down
electrons are reflected back to the left lead. Therefore, for the
energy in the direct subgaps opened by the Rashba spin-orbit
coupling, the spin polarization can reach a very large value as the
length of the nanoribbon increases, as shown in Fig. 3(c).

\begin{figure}[htb]
\centering
\includegraphics[scale=0.5,angle=0]{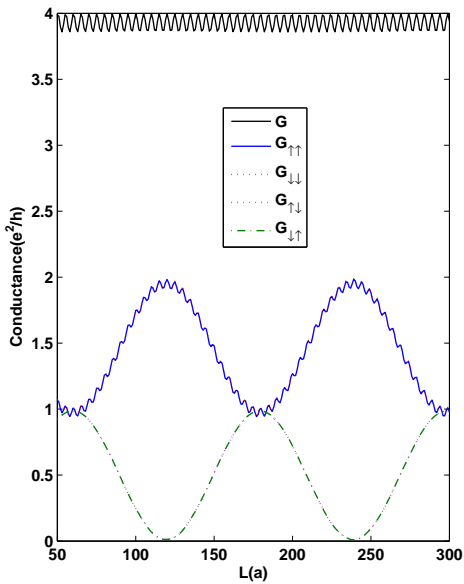}
\caption{(Color online) The total conductance $G$, spin-dependent conductance $G_{\sigma\sigma'}$ and spin polarization $P$ vs the length $L$ of the zigzag
silicene nanoribbon for $\lambda_{R}=0.07t$ and $E=-0.6t$.} \label{figfour}
\end{figure}

Fig. 4 shows the total conductance $G$ and spin-dependent
conductance $G_{\sigma\sigma'}$ versus the length of the zigzag
silicene nanoribbon $L$ for $E=-0.6t$, which correspond to the
``parallel crossing'', i.e., interplay between the traveling edge
and bulk states with the same direction of velocities. The largest
value of the total conductance is $4e^{2}/h$ because there are four
spin-dependent modes, including a spin-up mode, a spin-down mode,
and two spin-mixing modes, at the Fermi energy contributing to the
conductance. Due to the mismatch between the sample and the leads,
the total conductance shows resonant transmission properties. From
Fig. 4 we can see that
$G_{\uparrow\uparrow}=G_{\downarrow\downarrow}$ and
$G_{\uparrow\downarrow}=G_{\downarrow\uparrow}$, so the spin
polarization can not occur at the energy. However, all of the
spin-dependent conductances vary periodically as the length of the
zigzag silicene nanoribbon $L$ varies. This periodical variation of
the spin-dependent conductances originates from the spin mixing
induced by the interplay between the traveling edge and bulk states
with the same velocity. Due to the Rashba spin-orbit coupling, the
spin precesses in the nanoribbon at the energy. As the length
gradually increases, the amount of spin rolling from left to right
will change periodically. Therefore, the perfect spin modulation of
conductance occurs in the situation of parallel crossing.

Before arriving at the final summary, we emphasize that the above
phenomena are not a simple finite-size effect. With increasing
transverse size, there will be more bulk channels crossing the edge
channels, therefore there will be many parallel crossings (showing
periodic and spin-dependent conductances) and anti-parallel
crossings (showing spin polarization).

In summary, we study the interplay of the edge and bulk states
induced by the Rashba spin-orbit coupling in a zigzag silicene
nanoribbon in the presence of an external electric field. We find
that the interplay can be classified two types: (i) anti-parallel
crossing, the interplay between the edge and bulk states with
opposite velocities, which opens small and direct spin-dependent
subgaps; (ii) parallel crossing, the interplay between the edge and
bulk states with same velocity direction, which gives rise to the
significant anticrossing of the subbands. The spin-dependent
transport properties of the zigzag silicene nanoribbon are also
investigated by using nonequilibrium Green's function method. For
the former, a spin-polarized current in the nanoribbon can be
generated as the Fermi energy is in the direct spin-dependent
subgaps. While the later can give rise to the spin precession in the
nanoribbon. The interplay of the edge and bulk states induced by the
Rashba spin-orbit coupling in the zigzag silicene nanoribbon is
different from that of subbands in conventional semiconductor
nanoribbon, in which the Rashba spin-orbit coupling can only gives
rise to the spin precession.

This work was supported by National Natural Science Foundation of
China (Grant Nos. 11104059 and 61176089), Hebei province Natural
Science Foundation of China (Grant No. A2011208010), and
Postdoctoral Science Foundation of China (Grant No. 2012M510523).

\end{document}